\documentstyle[12pt,cite]{article}
\newcommand{\norm}[1]{{\protect\normalsize{#1}}}
\newcommand{\LAP}
{{\small E}\norm{N}{\large S}{\Large L}{\large A}\norm{P}{\small P}}

\newcommand{\be}{\begin{equation}}
\newcommand{\ee}{\end{equation}}
\newcommand{\bea}{\begin{eqnarray}}
\newcommand{\ena}{\end{eqnarray}}
\newcommand{\beano}{\begin{eqnarray*}}
\newcommand{\enano}{\end{eqnarray*}}
\newcommand{\sect}[1]{\setcounter{equation}{0}\section{#1}}
\newcommand{\vs}[1]{\rule[- #1 mm]{0mm}{#1 mm}}

\newcommand{\sm}[2]{\frac{\mbox{\footnotesize #1}\vs{-2}}
                   {\vs{-2}\mbox{\footnotesize #2}}}

\newcommand{\shalf}{\sm{1}{2}}

\newcommand{\var}{\varphi}
\newcommand{\vareps}{\varepsilon}
\newcommand{\overl}[1]{\overline{#1}}

\newcommand{\cd}{\mbox{$\cal{D}$}}
\newcommand{\cg}{{\mbox{$\cal{G}$}}}

\newcommand{\cp}{\mbox{$\cal{P}$}}

\newcommand{\cw}{\mbox{$\cal{W}$}}
\newcommand{\ca}{{\mbox{$\cal{A}$}}}
\newcommand{\co}{{\mbox{$\cal{O}$}}}

\newcommand{\undl}{\underline{\lambda}}

\newcommand{\prt}{\partial}
\newcommand{\eps}{\epsilon}
\newcommand{\und}[1]{\underline{#1}}

\newcommand{\om}{\omega}
\newcommand{\Om}{\Omega}
\newcommand{\lie}{\mbox{Lie}}
\newcommand{\tildom}{\tilde{\omega}}
\newcommand{\Tildom}{\tilde{\Omega}}
\newcommand{\barom}{\bar{\omega}}
\newcommand{\Barom}{\bar{\Omega}}
\newcommand{\Stop}{s^{top}}
\newcommand{\al}{\alpha}
\newcommand{\g}{\mbox{\scriptsize{g}}}
\newcommand{\m}{\mbox{\large{$m$}}}

\begin{document}
\renewcommand{\thefootnote}{\fnsymbol{footnote}}
\newpage
\pagestyle{empty}
\setcounter{page}{0}

\def\logolapin{
  \raisebox{-1.2cm}{\epsfbox{/lapphp8/keklapp/ragoucy/paper/enslapp.ps}}}
\def\logolight{{\bf {\large E}{\Large N}{\LARGE S}{\huge L}{\LARGE
        A}{\Large P}{\large P} }}
\def\logoenslapp{\logolight}
%
%
%
\hbox to \hsize{
\hss
\begin{minipage}{5.2cm}
  \begin{center}
    {\bf Groupe d'Annecy\\ \ \\
      Laboratoire d'Annecy-le-Vieux de Physique des Particules}
  \end{center}
\end{minipage}
\hfill
\logoenslapp
\hfill
\begin{minipage}{4.2cm}
  \begin{center}
    {\bf Groupe de Lyon\\ \ \\
      {\'E}cole Normale Sup{\'e}rieure de Lyon}
  \end{center}
\end{minipage}
\hss}

\vspace {.3cm}
\centerline{\rule{12cm}{.42mm}}

\vs{5}

\begin{center}

{\bf EXERCISES IN EQUIVARIANT COHOMOLOGY\footnote{Lectures given at
the NATO ASI, "Quantum fields and quantum space-time", Carg{\`e}se,\\
July 22-August 3, 1996.}
}\\[1cm]

\vs{5}

{\large R. Stora}\\[1cm]

{\em Laboratoire de Physique Th{\'e}orique }\LAP\footnote{URA 14-36
du CNRS, associ{\'e}e {\`a} l'Ecole Normale Sup{\'e}rieure de Lyon et {\`a}
l'Universit{\'e} de Savoie}\\
{\em LAPP, Chemin de Bellevue, BP 110\\ 
F-74941 Annecy-le-Vieux Cedex, France.\\[0.2cm]
and\\[0.2cm]
CERN, Theory Division, CH-1211 Gen{\`e}ve 23, Switzerland}\\

\end{center}
\vs{15}

\centerline{ {\bf Abstract}}

\indent

Equivariant cohomology is a mathematical framework particularly well 
adapted to a kinematical understanding of topological gauge theories
of the cohomological type. It also sheds some light on gauge
fixing, a necessary field theory operation connected with the non
compactness of the gauge group. The respective roles of fields and
observables are emphasized throughout.

\vs{5}

\rightline{\LAP-A-619/96}
\rightline{CERN-TH/96-279}
\rightline{October 1996}

\newpage
\pagestyle{plain}
\renewcommand{\thefootnote}{\arabic{footnote}}

\sect{INTRODUCTION}

\indent

Equivariant cohomology\cite{1}$^-$\cite{5} is at the core of the
geometrical interpretation of the topological -more precisely,
cohomological- field theories proposed by E. Witten in
1988\cite{6,7}. The corresponding mathematical equipment
also sheds some light on the gauge fixing procedure\cite{8}
familiar in the Lagrangian formulation of gauge theories.

The purpose of these notes is to update and, at times, provide
alternatives to the second part of the 1994 Les Houches lectures by
S.~Cordes, G.~Moore, S.~Rangoolam\cite{9}.

The first part is devoted to definitions and basic facts related to
equivariant cohomology. It relies on a systematic use of the Weil
algebra, which not only streamlines many of the basic calculations,
but is also of prime relevance in the geometrical framework for
cohomological field theories. This is the subject of section
2.

Section 3 gathers three exercises.

The first one consists of a quick derivation of the Matha{\"\i}
Quillen formulae\cite{3} which give special integral
representations of the Thom class of a vector bundle\cite{9}. It
is accompanied by another similar formula which implements "Cartan's theorem 3" described in
section 2\cite{1}.

The second one describes a general procedure to construct universal
observables pertaining to cohomological field theories.

The third one completes the construction of actions for
cohomological models as occuring in integral representations of
integrals over orbit spaces of cohomology classes of top dimension.
These actions involve, among others, gauge fixing terms whose
cohomological interpretaion is displayed as well. For the sake of
definiteness the Yang Mills case is thoroughly discussed.

\sect{EQUIVARIANT COHOMOLOGY.\\
DEFINITIONS AND ELEMENTARY PROPERTIES [1]-[4]}

\indent

{\bf Example 1}

Let $M$ be a smooth manifold, $\Omega^*(M)$ the differential forms
on $M, d_M$ the exterior differential.

Let $\cg$ be a connected Lie group, Lie $\cg$ its Lie algebra,
acting smoothly on $M$. $\lambda \in$ Lie $\cg$ is represented by a
vector field on $M,\underline{\lambda}$. For each $\om \in \Om^*(M)$, define
\bea
i_M(\lambda)\om &=& i(\undl)\om \nonumber\\
\ell_M(\lambda) \om &=& \ell (\undl) \om
\label{eq:1}
\ena
where $i(\undl)$ denotes the interior product by the vector field
$\undl$, and $\ell (\undl)$, the corresponding Lie derivative:
\be
\begin{array}{rcll}
\ell_M (\lambda) &=& [i_M(\lambda),d_M]_+ & \\
{[}i_M(\lambda), i_M(\lambda ')]_+ &=& 0 &  \\
{[} \ell_M (\lambda), i_M (\lambda ')]_- &=& i_M([\lambda, \lambda
']) & \ \ \; \; \lambda, \lambda ' \in 
\mbox{Lie} \cg \\
{[} \ell_M(\lambda), \ell_M(\lambda ')]_- &=& \ell_M
([\lambda,\lambda ']) & 
\end{array}
\label{eq:2}
\ee
where $[\lambda, \lambda ']$ is the bracket in Lie $\cg$ (recall
$[\undl, \undl ']_{Lie} = [\underline{\lambda, \lambda '}]$).

Forms $\om$ such that $i_M(\lambda)\om=0 \; \forall \lambda \in$ Lie
$\cg$ are called horizontal.

Forms $\om$ such that $\ell_M (\lambda) \om=0 \; \forall \lambda \in$
Lie $\cg$ are called invariant.

Forms which are both horizontal and invariant are called basic.

Basic cohomology is the de Rham cohomology restricted to basic forms
which obviously span a subcomplex of the de Rham complex. 

\indent

{\bf Remark}

The term basic refers to the following situation: if $M$ is locally
of the form $X \times \cg$ with local coordinates $x \in X, y
\in \cg$, one can write locally
\be
\om(x, y, dx, dy) = \Sigma \ \om_{MN} (x,y) \ dx^M \ dy^N
\label{eq:3}
\ee
where $M$ and $N$ are multi indices.

Horizontality implies $N=0$.

Invariance implies $\om_{MO}$ is independent of $y$.

Thus
\be
\om_{basic} (x,y \ dx, dy) = \Sigma \ \om_M (x) \ dx^M
\label{eq:4}
\ee
i.e. can be identified with a form on the base $X$.

\indent

{\bf Example 2}

The Weil algebra of $\cg : \cw(\cg)$. One defines $\cw(\cg)$ as a
graded commutative differential algebra: as a vector space
\be
\cw(\cg) = \wedge (\lie \ \cg)^* \otimes S\left( (\lie \
\cg)^*\right) 
\ee
where $\wedge (\lie \ \cg)^*$ is the Grassmann algebra of $(\lie
\ \cg)^*$ and $S((\lie \ \cg)^*)$ is the symmetric tensor algebra of
$(\lie \ \cg)^* \cdot \cw(\cg)$ is generated by Lie $\cg$ valued
generators $\om$ (for $\wedge (\lie \ \cg)^*)$, assigned grading 1,
$\Om$ (for $S((\lie \ \cg)^*)$), assigned grading 2. One defines the
differential $d_W$ by
\bea
d_W \ \om &=& \Om - \shalf [\om, \om] \nonumber \\
d_W \ \Om &=& - [\om, \Om] \cdot
\label{eq:5}
\ena

\indent

{\bf Remark}

Eqs (\ref{eq:5}) are an algebraic abstraction of the relationship
between the "curvature" $\Om$ of a connection $\om$ on a principal
$\cg$ bundle, and the latter: the first one defines the curvature;
the second one is the Bianchi identity $"d^2_W =0"$, via the Jacobi
identity $[\om, [\om, \om]]=0$.

\indent

An action of Lie $\cg$ on $\cw(\cg)$ is defined by
\be
\begin{array}{ll}
i_W (\lambda) \om = \lambda & i_W (\lambda) \Om =0 \\
\ell_W (\lambda) \om = [\lambda, \om] & \ell_W (\lambda) \Om =
[\lambda, \Om].
\label{eq:6}
\end{array}
\ee
$i_W(\lambda), \ell_W(\lambda), d_W$ fulfill the analogs of
Eqs(\ref{eq:2}) with the subscript $W$ replacing the subscript $M$.

\indent

The cohomology of $\cw(\cg)$ is easily seen to reduce to the scalars.
 Introduce for instance the homotopy\cite{10}
\be
k = \int^1_0 k_t \varphi_t
\label{eq:7}
\ee
where
\be
\varphi_t = \om_t \ \frac{\prt}{\prt \om} + \Om_t \frac{\prt}{\prt
\Om}
\label{eq:8}
\ee
with, e.g.
\be
\om_t = t \om \; \  \Om_t = d_W \om_t + \shalf [\om_t, \om_t]
\label{eq:9}
\ee
and
\be
k_t \ \Om_t = d_t \om_t \; \; k_t \om_t =0 \cdot
\label{eq:10}
\ee
One easily proves
\be
k_t \ d_W + d_W \ k_t = d_t = dt \ \frac{\prt}{\prt t} \cdot
\label{eq:11}
\ee

The basic cohomology, on the other hand consists of all the $\cg$
invariant polynomials in $\Om$.

\indent

{\bf Example 3}

$(E, d_E)$ is a graded commutative differential algebra with an
action $i_E(\lambda) , \ell_E(\lambda)$ of Lie $\cg$ fulfilling
relations (\ref{eq:2}) (with the subscript $M$ replaced by $E$).
Examples 1 and 2 are particular cases of this. 

\indent

{\bf Equivariant cohomology}\cite{1,4}.

\underline{Def.}: The equivariant cohomology of
$(E,d_E),(i_E(\lambda), \ell_E(\lambda))$ is defined as the basic
cohomology of $\left( \cw(\cg) \otimes E, D_W = d_W + d_E \right),
I_W \left( (\lambda) = i_W(\lambda) + i_E(\lambda), L_W(\lambda) =
\ell_W(\lambda) + \ell_E(\lambda) \right)$.

This is the so called Weil model for equivariant cohomology.

As we shall see later, the intermediate model to be presently
defined is of substantial interest in practical computations. The
following construction is due  to J.~Kalkman\cite{4}. It owes its
simplicity to the systematic use of the Weil algebra and \underline{not} 
of concrete connections.

Under the automorphism
\be
x \rightarrow e^{i_E(\om)} x
\label{eq:12}
\ee
the differential $D_W$ goes into
\be
D_{int} = d_W + d_E + \ell_E(\om) - i_E (\Om)
\label{eq:13}
\ee
and the operation is transformed into
\bea
I_{int} (\lambda) &=& i_W (\lambda)
\label{eq:14}\\
L_{int} (\lambda) &=& \ell_W (\lambda) + \ell_E (\lambda) \cdot
\label{eq:15}
\ena

The most efficient way to do these calculations is to look at the
family $D_t, I_t(\lambda), L_t(\lambda)$ conjugate to the initial
data through the automorphism $x \rightarrow e^{t \ i_E(\om)}x$. See also
ref [5].  As we shall see later, the efficiency of this
model is due to the simplicity of $I_{int} (\lambda) = i_W (\lambda)$.

\indent

{\bf Remark}

If $\om, \Om$ were replaced by a connection $\tilde{\om}_E$ on $E$
(whose definition will be given later) and its curvature
$\tilde{\Om}_E$, the whole construction would be spoiled !

It is fairly easy to see that the basic cohomology, in this scheme,
is realized on the $\cg$ invariant elements of $S((\lie \cg)^*)
\otimes E$ endowned with the Cartan differential $D_C = d_E - i_E
(\Om). \ (D^2_C = \ell_E (\Om)$ vanishes on $\cg$ invariant
elements). The Cartan model is the most popular, although not
always the most convenient.

\indent

{\bf Remarks}

As mentioned earlier, basic cohomology is related to the
"cohomology of the base" when the latter exists e.g., in the case of
the action of $\cg$ on a manifold $M$, when $M/\cg$ exists as a
manifold, i.e. when $M$ 
is a principal $\cg$ bundle. This situation is characterized by the 
existence of $\cg$
connections on $M$, i.e. Lie $\cg$ valued one forms $\tilde{\om}$
such that $i_M(\lambda) \tilde{\om} = \lambda \; \; \ell_M (\lambda)
\tilde{\om} = [ \lambda, \tildom ]$. This generalizes to arbitrary
commutative differential algebras $(E,d_E)$ for which a connection
$\tildom$ is defined as a Lie $\cg$ valued element of grading one
such that $i_E (\lambda) \tildom = \lambda$ $\; \ell_E (\lambda)
\tildom = [ \lambda, \tildom ]$. In fact, in this case, equivariant
cohomology is isomorphic with basic cohomology. This is Cartan's
"theorem 3"\cite{1,5}, which implies in particular that
this holds independently of the choice of a connection $\tildom$.

The proof is easy. First a basic cohomology class is obviously an
equivariant cohomology class. Conversely, given an equivariant
cohomology class with representative $P(\om, \Om,x), x \in E$ we
have, using the homotopy\cite{10,11} of the Weil
algebra \be
P(\om, \Om, x) - P( \tildom, \Tildom,x) = \int^1_0 k_t (D_W P)
(\om_t, \Om_t, x) + D_W(k_t P)(\om_t \Om_t,x)
\label{eq:16}
\ee
where now 
\be
D_W = d_W + d_E
\label{eq:17}
\ee
\be
\om_t = t\om +(1-t)\tildom .
\label{eq:18}
\ee
The conclusion follows from 
\be
I_W(\lambda) \om_t = [i_W (\lambda) +
i_E(\lambda)] \om_t = \lambda \cdot
\label{eq:19}
\ee
\be
L_W(\lambda) \om_t = [\ell_W(\lambda) + \ell_E(\lambda) ] \om_t =
[\lambda, \om_t]
\label{eq:20}
\ee

Before concluding this section, we should mention that we have
described equivariant cohomology at the Lie algebra level as it was
introduced by H. Cartan\cite{1}. It only co{\"\i}ncides with that
defined in terms of the classifying space of $\cg$ when $\cg$ is
compact\cite{1}. Although global phenomena have to be kept in
mind we shall stick here to the infinitesimal description in view of the
forthcoming applications to cohomological field theory models.

\sect{THREE EXERCISES}

The main leitmotiv of this section is the integral representation of
the Dirac current located at the origin of a vector space $V$:
\be
\Om_0 \equiv \delta(v) \wedge dv = N_0 \int {\cd} b {\cd}
\bar{\om} \; \; e^{-i<b,v> + <\bar{\om},dv>}
\label{eq:21}
\ee
where $v$ denotes a set of coordinates in $V, \wedge dv$ denotes
the corresponding volume form, $b$ denotes a set of dual coordinates in
the dual $V^*, \bar{\om}$ denotes the corresponding set of
generators of $\wedge V^*, \cd b = \wedge db$, $\cd \bar{\om}$ is
the volume element for Berezin integration, $< \ ,\ >$ denotes the
duality pairing, $N_0$ is a normalization constant only depending on
$|V|$, the dimension of $V$.

This Dirac current is invariant under linear changes of coordinates,
being a product of a bosonic and a fermionic $\delta$ function.

Another suggestive representation is:
\be
\Om_0  \equiv \delta(v) \wedge dv = N_0 \int \cd b \cd \bar{\om}
\; \; e^{s(<\bar{\om}, v>)}
\label{eq:22}
\ee
where $s$ is defined as 
\be
sv = dv \; \; \; \; \; s \ dv =0
\label{eq:23}
\ee
and is extended to the integration variables by
\be
s \barom = ib \; \; \; \; sb=0 \cdot
\label{eq:24}
\ee

Other representatives of the top cohomology of $V$ with compact
supports are obtained by perturbing the function under the $s$
operation
\be
\Om = N_0 \int {\cd} \barom  {\cd}
b \; \; e^{-s(<\barom ,v> - i<\barom , \var (b)>)}
\label{eq:25}
\ee
for $\var$'s such that $<b, \var(b)>$ is positive at infinity and
$\var(0)$ is bounded.

\indent

{\bf Exercise 1:} The Thom class of a vector
bundle\cite{3,4,9}
and Cartan's theorem 3\cite{1}.

Let $E(M,V)$ be a vector bundle with fiber $V$ associated to a
principal $\cg$ bundle $P(M,\cg)$:
\be
E(M,V) = P(M,\cg) \times_{\cg} V
\label{eq:26}
\ee
for some representation $R_V$ of $\cg$ in $V$ with differential
$t_V$ representing Lie \cg.

The Dirac current on $V$ defines a distributional form on $E$
representing the "Poincar{\'e} dual"\cite{9} of the section $V=0$
of $E$, diffeomorphic to $M$. Smooth representatives of this cohomology
class with fast decrease along the fibers have been constructed by
V.~Matha{\"\i} and D.~Quillen\cite{3}.

This can be compactly described as follows: construct in the
intermediate model the equivariant class which extends a gaussian
thickening of the Dirac current, namely $e^{-\frac{(v,v)}{2}} \wedge
dv$, where $(v,v)$ is a $\cg$ invariant form on $V$. This can be
written as
\be
\Theta_{int} = N_0 \int \cd b \cd \barom \; \;e^{s^{top} (<\barom,v>
- i (\barom, b)^*)}
\label{eq:27}
\ee
where
\bea
\Stop \ V &=& \psi_V + t_V(\om) v \nonumber \\
\Stop \ \psi_V &=& t_V(\om) \psi - t_V (\Om) V \; \; \; \ \ \; \psi_v
= dv \nonumber \\
\Stop \ \om &=& \Om - \shalf [\om, \om] \nonumber \\
\Stop \ \Om &=& - [\om, \Om] \nonumber \\
\Stop \ \barom &=& ib - \barom t_V (\om) \nonumber \\
\Stop \ ib &=& ib  \ t_V (\om) - \barom t_V (\Om)
\label{eq:28}
\ena
and $(\ ,\ )^*$ is the invariant form on $V^*$ which yields $(\ , \
)$ on $V$. In the Weil model, one only needs to replace $\psi_V =
dv$ by $\psi_V = dv - t_V(\om) v$.

The extension of $\Stop$ to the integration variables not only
allows a compact writing but also provides easy proofs that $\Theta$
is equivariantly closed. $\Theta$ is called the universal Thom class
of $E$. Replacing $\om$ by $\tildom$, a connection on $P(M,\cg)$ and
$\Om$ by $\Tildom$, the curvature of $\tildom$ provides a globally
defined form on $E$, the Thom class of $E$. By construction, smooth
deformations of the function under the $\Stop$ sign leave one in
the same cohomology class. So does a variation of $\tildom$.
Replacing $v$ by $v(x), x \in M$, a smooth section of $E$ produces
a cohomology class of $M$ located at the zeroes of that section, the
so-called Euler class of $E$, independent of the choice of the
section $v=v(x)$.

Another construction of the same type provides the identity which
implements Cartan's theorem 3:
\be
\int \cd \om \cd \Om \ \delta(\om - \tildom) \ \delta (\Om - \Tildom)
=1 \cdot \label{eq:29}
\ee

Introducing integral representations for both the fermionic and the
bosonic $\delta$ functions, we get
\be
\int \cd \om \cd \Om \cd \barom \cd \Barom \; \; e^{\barom (\om -
\tildom) + i \Barom (\Om - \Barom)} =1 \cdot
\label{eq:30}
\ee

Extending $\Stop$ as usual to $P(M, \cg)$ and, to the integration
variables, by
\bea
s \; i \Barom &=& \barom + [\om , i \Barom] \nonumber \\
s \barom &=& - [\Om , i \Barom] + [ \om, \barom] \cdot
\label{eq:31}
\ena

This can be written
\be
\int \cd \om \cd \Om \cd \barom \cd \Barom \; e^{\Stop \; i 
\Barom (\om - \barom)} =1 \cdot
\label{eq:32}
\ee

One may define $\Stop$ on $P$ by
\bea
\Stop p &=& \psi_p + \co_p (\om, p) \nonumber \\
\Stop \psi_p &=& - \co_p (\Om , p) - \frac{\prt \theta p}{\prt
p} (\om, p) \cdot \psi_p
\label{eq:33}
\ena
where $p$ denotes a set of coordinates on $P$. This implies
\bea
\Stop \tildom &=& \Tildom - \shalf [ \tildom , \tildom ] \nonumber
\\
\Stop \Tildom &=& - [ \tildom, \Tildom ] \cdot
\ena

If one defines $\tildom$ by
\be
H(p) \psi_p = 0
\label{eq:35}
\ee
where
\be
\psi_p = dp + \co (\tildom , p)
\label{eq:36}
\ee
and $H(p) \psi_p$ has values in Lie $\cg$, one also has
\be
\int \cd \om \cd \Om \cd \barom \cd \Barom \; \; e^{\Stop (i \Barom
H(p) \psi_p)} =1 \cdot
\label{eq:37}
\ee

\indent

{\bf Exercise 2:} Universal observables for cohomological models as
equivariant characteristic classes\cite{12,13,5}
.

We return to the situation where the manifold $M$ is smoothly acted
on by the connected Lie group $\cg$. Assume the action of $\cg$ can be
lifted to a $K$ principal bundle $P(M,K)$ over $M$ on which
there is a $\cg$ invariant  $K$ connection $\Gamma$.

Define the equivariant curvature in the intermediate model
\be
R^{eq}_{int} (\Gamma) = D_{int} \ \Gamma + \shalf [\Gamma, \Gamma ]
= R(\Gamma) - i_P (\Om) \Gamma
\label{eq:38}
\ee
and the corresponding $K$-characteristic classes, $\cp_K
(R^{eq}_{int} (\Gamma))$ where the $\cp_K$'s are $K$ invariant
symmetric polynomials on Lie $K$. It is easy to prove that these
define equivariant classes of $M$ independent of the choice of
$\Gamma$. These can be written down in the Weil model upon operating
with $e^{i_P(\om)}$.

It turns out that these classes exhaust the examples of observables
known for cohomological gauge models\cite{6}.

The case of the topological Yang Mills theory in four dimensions is
well known. Let $\ca$ be the space of connections on $P(M,G)$. On
$P(M,G) \times \ca$, one takes the invariant $\cg$ connection
$\underline{a}$ (considered as a one form on $P$ and a zero form -coordinate function- on $\ca$). The equivariant curvature
is, in the intermediate model
\bea
R^{eq}_{int} (\underline{a}) &=& F(a) + \delta a + \Om \nonumber \\
&=& F(a) + \psi_{int} + \Om \cdot
\label{eq:39}
\ena

In the Weil model $\underline{a}$ is transformed into
$\underline{a} + \om$
\be
R^{eq}_{w} (\underline{a} + \om) = F(a) + \psi_w + \Om \cdot
\label{eq:40}
\ee

Taking a $G$ invariant symmetric polynomial on Lie $G \ P_G$ and
expanding $P_G (R^{eq}_w)$ into monomials $P^{\g}_{4-\g}$ of
bidegree 4-g on $M,$ g on $\ca$, yields observables upon
integration over a cycle $\gamma_{4-\g}$ of dimension 4-g in $M$:
\be
\co^{\g}_{4-{\g}} (\gamma_{4-{\g}}) = \int_{\gamma_{4-{\g}}}
P^{\g}_{4-{\g}} \label{eq:41}
\ee
whose cohomology class only depends on the homology class of
$\gamma_{4-{\g}}$.

\indent

{\bf Exercise 3:} Cohomological models and integral representations
of orbit space integrals of top cohomology classes: Yang Mills
theories\cite{6,14,16,17,21}.

The problem of integrating basic cohomology classes over orbit space
is inherent to the present field theory formulations of gauge
theories. Dynamical gauge theories are defined via a $\cg$ invariant
top form on a principal $\cg$ bundle. For instance, Yang Mills
theories are defined on $\ca = P(\ca/\cg,\cg)$ the space of
connections $\underline{a}$ on a principal bundle $P(M,G)$ where $G$
is a compact Lie group, and $\cg$ its gauge group, which is  non
compact. The Yang Mills form 
\be
\Om_{YM} = e^{-S_{inv}(a)} \cd a
\label{eq:42}
\ee
defines the dynamics, but is not integrable in the fiber direction.
One wishes to integrate gauge invariant observables $\co_{inv}(a)
\Om_{YM}$. In most textbooks, the Faddeev Popov\cite{8} gauge
fixing procedure is presented by factoring out the volume of the
gauge group. In J.~Zinn-Justin's book\cite{18} on the other hand
this is achieved by "integrating over the fiber" a route we shall
now follow. Given a \cg invariant volume form $\mu$ on $\cg$ and its
dual $\tilde{\mu}_{\cg}$ on Lie \cg normalized so that  \be
<\mu_{\cg},
\tilde{\mu}_{\cg}>=1,
\label{eq:43}
\ee
one can construct the Ruelle Sullivan\cite{19} closed basic form
\be
\Om_{RS} = i(\underline{\tilde{\mu}}_{\cg})\Om_{YM}
\label{eq:44}
\ee
where $\underline{\tilde{\mu}}_\cg$ is obtained by substituting
for each element $X_\al$ of Lie $\cg$ the corresponding fundamental
vector field $\underline{X}_\al$ on $\ca$. Closedness is a consequence of
both the closedness of $\Om_{YM}$ and of its invariance.
Horizontality is obvious as well as invariance.

Both $\co_{inv}(a)$ and $\Om_{RS}$ define objects on $\ca /
\cg$, respectively a function and a top form denoted $\tilde{\co} (a),
\Tildom_{RS}$, and one wishes to integrate $\tilde{\co}(a)
\Tildom_{RS}$ over $\ca / \cg$. Choosing a locally finite covering
$U_i$ of $\ca / \cg$ and a partition of unity $\theta_i(\dot{a})$
subordinate to it as well as local sections $\sigma_i$ defined by
local equations 
\be
g_i(a)=0
\label{eq:45}
\ee
above each $U_i$, we may write
\bea
<\co > &=& \int_{\ca / \cg} \tilde{\co}(a) \Tildom_{RS} = \sum_i
\int_{\ca / \cg} \theta_i (\dot{a}) \co(a) \Om_{RS} \int_{fiber}
\delta (g_i) \wedge \delta g_i \nonumber \\
&=& \int_{\ca} \sum_i \theta_i (\dot{a}) \delta (g_i) \wedge \delta
g_i \co(a) \Om_{RS} \cdot
\label{eq:46}
\ena
Gauge independence is due to the closedness and basicity of
$\theta_i(\dot{a}) \co (a) \Om_{RS}$: at the infinitesimal level
$\delta(g_i) \wedge \delta g_i$ varies by $\ell(\hat{\om}) [\delta
(g_i) \wedge \delta g_i]$ where $\hat{\om}$ is the vertical vector
field defined by 
\be
\ell_{\hat{\om}} g_i = \Delta g_i
\label{eq:46bis}
\ee
where $\Delta$ denotes the infinitesimal change of section. The
differential form
\be
\gamma = \sum_i \theta_i (\dot{a}) \delta (g_i) \wedge \delta g_i
\label{eq:47}
\ee
will be called the gauge fixing form. It has the following
property: its projection $\hat{\gamma}$ on any fiber\cite{20},
i.e., the representative of its class modulo the ideal generated by
horizontal forms of strictly positive degree is a top class of $\cg$
whose integral is one: \be
\int_{Fiber} \hat{\gamma} =1
\ee

It therefore projects on the top cohomology class of $\cg$ with
compact supports (or fast decrease), normalized to 1. Since
$\Om_{RS}$ is a top basic form, it is clear that only the fiber
projection of $\gamma$ matters. $\gamma$ may be considered as the
Poincar{\'e} dual of a section which does not exist.

So, we may as well represent our integral as
\be
<\co > = \int_{\ca / \cg} \tilde{\co} (a) \Tildom_{RS} =
\int_\ca \gamma \co(a) \Om_{RS} = \int_\ca \hat{\gamma} \co(a) \Om_{RS}
\label{eq:47bis}
\ee
where $\hat{\gamma}$ can be defined by choosing a connection
$\tildom$:
\be
\hat{\gamma} = \gamma (a) \mu_{\cg} (\tildom)
\label{eq:48}
\ee
so that
\be
< \co > = \int_{\ca / \cg} \tilde{\co} (a) \Tildom_{RS} =
\int_\ca \gamma(a) \co(a) \Om_{YM} \cdot
\label{eq:49}
\ee

This can be algebraized as
\be
< \co > = \int \tilde{\mu}_\cg (\cd \om) \hat{\gamma} (a, \om)
\co(a) \Om_{YM}.
\label{eq:50}
\ee
where a Berezin integral over Lie $\cg$ is included, and
\be
\hat{\gamma} (a,\om) = \gamma (a) \mu_\cg (\om) \cdot
\label{eq:51}
\ee

By construction
\be
s \hat{\gamma} (a, \om) =0
\label{eq:52}
\ee
with
\bea
sa &=& \ell_\om a \nonumber \\
s \om &=& - \shalf [\om, \om ]
\label{eq:53}
\ena
and $\hat{\gamma}$ is ambiguous up to a coboundary:
\be
\hat{\gamma} \rightarrow \hat{\gamma} + s \chi
\label{eq:54}
\ee
which is a consequence of the fact that two top classes of $\cg$
with compact support which integrate to 1 differ by a coboundary.
These constructions can be taken as a basis for the geometrical
origin of the Slavnov symmetry. When $\gamma$ is constructed by
patching up local sections, as in eq. (\ref{eq:47}), one has the
well known formulae
\bea
\hat{\gamma}_i (a,\om) &=& \delta (g_i) \wedge sg_i \nonumber \\
&=& \int \cd b \cd \barom \; \; e^{i<b,g_i> - <\barom , sg_i>} \cdot
\label{eq:55}
\ena

Extending the operation $s$ to the integration variables
\bea
s \barom &=& ib \nonumber \\
sb &=& 0 \cdot
\label{eq:56}
\ena

This can be rewritten
\be
\hat{\gamma}_i (a, \om) = \int \cd b \cd \barom \; \; e^{s(<\barom,
g_i>)} \cdot \label{eq:57}
\ee

These formulae are valid as long as the manifolds $g_i=0$ stay
transverse to the fibers. The Faddeev Popov operator $m_i$ defined
by
\be
s g_i = \m_i \om
\label{eq:58}
\ee
is then invertible. The $\delta$ current can be smoothed out by
adding a term $<i \barom, \var (b)>$ under the $s$ operation where
$\var$ is such that $<b, \var (b)>$ is positive at infinity. This
suggests a class of gauge fixing forms
\be
\hat{\gamma} (a,\om) = \int \cd b \cd \barom \; \; e^{s(<\barom,
g(a,\dot{a})> + i<\barom, \var (b,\dot{a})>)}
\label{eq:59}
\ee
where $g(a,\dot{a})$ and $\var(b,\dot{a})$ involve an additional
orbit space dependence such that $m(a,\dot{a})$ remains invertible
and $\gamma(a,\om)$ has compact support or fast decrease. The
inclusion of such dependences is liable to spoil renormalizability
or locality or both.

The situation with topological theories is slightly more involved
and goes through  several steps.

Orbit space is restricted to the finite dimensional manifold of
gauge equivalence classes of solutions of a \cg invariant system of
equations $\vareps$ e.g. $ F^- \equiv F - * F = 0$. This restriction
is obtained via the insertion of the corresponding Matha{\"\i} Quillen
form $\Theta_\vareps$. One now considers an observable of degree the
dimensionality $d$ of this submanifold e.g. 
\be
\co_d = \prod_{{\Sigma \g_i=d}} \co^{\g_i} \cdot
\label{eq:60}
\ee

Choosing a connection $\tildom$ with curvature $\Tildom$ transforms
$\Theta_\vareps$ into $\tilde{\Theta}_\vareps$, $\co_d$ into
$\tilde{\co}_d$. In particular $\psi = \delta a + D_a \om$ is
transformed into $\tilde{\psi} = \delta a + D_a \tilde{\om}$.
$\tilde{\co}_d \tilde{\Theta}_\vareps$ defines a basic form which
can be integrated over $\ca / \cg$ after choosing local sections and a
partition of unity:
\bea
< \co_{|\vareps}> &=& \int_{\ca / \cg} \sum_i \theta_i(\dot{a})
\; \sigma^*_i (\tilde{\co}_d \tilde{\Theta_\vareps}) \nonumber \\
&=& \int \sum_i \theta_i (\dot{a}) \cd \om \cd \Om (\co_d
\Theta_\vareps)(a, \Tildom, \tilde{\psi}) \nonumber\\
&& \delta (\om -
\tildom) \delta (\Om - \Tildom) \delta (g_i) \wedge \frac{\delta g_i}{\delta a} D_a
\tildom
\label{eq:61}
\ena
where $\tilde{\psi}= \delta a + D_a \tildom$ and $\Tildom$ is
quadratic in $\tilde{\psi}$. In the last factor one can write
indifferently $\frac{\delta g_i}{\delta a} \delta a= \frac{\delta g_i}
{\delta a} (\tilde{\psi} - \cd_a \tildom)$ or $\frac{\delta
g_i}{\delta a} D_a \tildom)= \m_i \tildom$ because $(\co_d
\Theta_\eps)$ has maximum degree in $\tilde{\psi}$. Using the
$\delta (\om - \tildom)$ factor we can replace everywhere
$\tilde{\psi}$ by $\psi = \delta a - D_a \om$ and, in the last
factor $\tildom$ by $\om$. Expressing also $\delta (\om-\tildom)$ in
terms of $\psi = \delta a - D_a \om$, we can rewrite
\bea
<\co_{|\vareps}> &=& \int \sum_i \theta_i (\dot{a}) \cd \om \cd
\Om (\co_d \Theta_\vareps) (a, \Tildom(\psi), \psi) \nonumber \\
&& \delta \left( (\om - \tildom (\psi) \right) \delta \left( \om -
\tildom (\psi) \right) \nonumber \\
&& \cd \psi \delta \left( \psi -
(\delta a + D_a \om) \right) \delta (g_i) \wedge \m_i \om \cdot
\label{eq:62}
\ena
Using now the fact that $(\co_d
\Theta_\vareps)(a,\Tildom(\psi), \psi)\delta(\om-\tildom)(\psi))$
is of maximum degree in $\psi$ and also that $\wedge m_i \om$ is of
maximum degree in $\om \ \delta (\psi - (\delta a + D_a\om))$ can
be replaced by $\delta (\delta a)=\cd a$. We end up with
\[
<\co_{|\vareps}> \int \sum_i \theta_i (\dot{a}) \cd \om \cd \Om \cd
\psi \cd a (\co_d \Theta_\vareps) (a,\Om,\psi) 
\]
\be
\delta \left( (\om - \tildom)(\psi) \delta \left( \Om - \Tildom
(\psi) \right) \right) \delta (g_i) \wedge \m_i \om
\label{eq:63}
\ee

As shown before, $\co_\vareps \delta ((\om - \tildom) (\psi))
\delta (\Om - \Tildom (\psi))$ can be expressed in terms of Fourier
transforms involving pairs of fermionic and bosonic integration
variables. In the present example, with 
\be
\tildom = - \frac{1}{D^*_a
D_a} D^*_a \delta a
\label{eq:64}
\ee
(which excludes reducible connections),
\bea
&&\Theta_\vareps \delta \left( (\om - \tildom)(\psi)\right) \delta
\left( \Om - \Tildom (\psi) \right) = \int \cd \barom^- \cd b^- \cd
\barom \cd \Barom \nonumber \\
&& e^{\Stop(\barom^- F^- + \Barom D^* \psi)} \nonumber \\
&&= e^{ib^-F^-+\barom^- (D \psi)^- + \barom D^* \psi + i \Barom
(D^* D \Om + [\psi^+, \psi])} \cdot
\label{eq:65}
\ena

The expression under the exponential can be supplemented by a term
consistent with power counting $tr [ \Barom, \Om] \barom$ yielding
in the action the sum of two terms : $tr \{ [ \Barom, \Om]^2 + [
\barom, \barom]\Om \}$.

The manipulations described above provide a bridge
between the interpretation of L.~Baulieu, I.M.~Singer
\cite{14,15} and that of M.F.~Atiyah,
L.~Jeffrey\cite{17}. They may seem fairly arbitrary and non
unique. They are geared towards field theory integral representations of orbit
space integrals whose ultraviolet singular behaviours have to be
controlled in an algebraic manner. In that respect, the gauge fixing
factor lacks an algebraic characterization which extends in a
satisfactory way that of the first part: the full action has the
form
\be
S(a, \psi , \om, \Om ; \barom^- b^- \barom \Barom) = \Stop \chi +
S_{gf}
\label{eq:66}
\ee

Let us recall
\bea
\Stop a &=& \psi - D_a \om \nonumber \\
\Stop \psi &=& [\om, \psi] - D_a \Om \nonumber \\
\Stop \om &=& \Om - \shalf [\om, \om] \nonumber \\
\Stop \Om &=& -  [\om, \Om] \nonumber \\
\Stop \barom^- &=& ib^- +  [\om, \barom^-] \nonumber \\
\Stop ib^- &=& [\om, ib^-] - [\Om, \barom^-] \nonumber \\
\Stop i\Barom &=& \barom + [\om, i\Barom] \nonumber \\
\Stop \barom &=& -[\Om , i \Barom] + [\om, \barom] \nonumber \\
I(\lambda) \om &=& \lambda \; \; \; \; I(\lambda) \mbox{other} =0 \;
\; \; \; \lambda  \in Lie \cg \nonumber\\
 L(\lambda) &=& [I(\lambda) , \Stop]_+
 \label{eq:67}
 \ena
 
 $\chi$ is basic:
 \be
 I(\lambda) \chi = L(\lambda) \chi =0 \; ; \; \lambda \in Lie \cg \cdot
 \label{eq:68}
 \ee
 
 In order to extend this structure to $S_{gf}$, we replace the
 basicity of $\chi$ by an equivalent condition. Define\cite{16}
 \be
 W =L(\underline{\lambda}) + I(\underline{\mu})
 \label{eq:69}
 \ee
 where $\underline{\lambda}$ generates $\wedge (Lie \cg)^*$ and
 $\underline{\mu}$ generates $S(Lie \cg^*)$. In particular
 \bea
 W \underline{\lambda} &=& - \shalf [ \underline{\lambda}, \underline{\lambda}
 ]_+ \nonumber \\
  W \underline{\mu} &=& - [ \underline{\lambda},
  \underline{\mu}  ]_- \cdot
  \label{eq:70}
  \ena
  
  By construction, $W^2 =0$.
  
  If one extends $\Stop$ to the algebra generated by $\und{\lambda},
  \und{\mu}$ according to
  \bea
  \Stop \und{\lambda} &=& \und{\mu} \nonumber \\
  \Stop \und{\mu} &=& 0
  \label{eq:71}
  \ena
  we have
  \be
  [ \Stop , w ] =0 \cdot
  \label{eq:72}
  \ee
  
  From now on, we shall omit the underlining of $\und{\lambda}$ and
  $\und{\mu}$ which will denote the Faddeev Popov ghosts of the
  graded Lie algebra generated by $L(\cdot), I(\cdot)$. It is clear
  that,   restricted to $\lambda, \mu$ independent elements, basicity is
  equivalent to $W$ invariance. Thus,
  \be
  W \chi = 0 \cdot
  \label{eq:73}
  \ee
  
  We now introduce Lagrange multipliers and antighosts
  $(\overl{\lambda}, \ell) (\overl{\mu}, m)$ associated to $\lambda,
  \mu$. We extend $W$ by
  \bea
  W \overl{\lambda} &=& \ell \nonumber \\
  W \ell &=& 0 \nonumber \\
  W \bar{\mu} &=& m \nonumber \\
  W m &=& 0 
  \label{eq:74}
  \ena
  $\Stop$ has to be extended in order to preserve its anticommutation
  with $W$:
  \be
  \begin{array}{ll}
  \Stop \bar{\mu} = \bar{\lambda} & \Stop \bar{\lambda} =0 \\
  \Stop m =-\ell & \Stop \ell =0 \cdot
  \end{array}
  \label{eq:75}
  \ee
  
  We now have a bigrading relative to the two differentials $\Stop,
  W$, as follows 
  \be
  \begin{array}{rccccccccccccc}
  &\Stop & W & a,&\psi,&\om,&\Om &etc. & \lambda & \mu &
  \bar{\lambda}   &   \bar{\mu} & \ell & m \\
 s \; \; \mbox{grading}  & 1&0&0&1&1&2&&0&1&0&-1&0&-1\\
 W \; \; \mbox{grading}  & 0&1&0&0&0&0&&1&1&-1&-1&0&0\\
 \label{eq:76}
\end{array}
\ee
 
 Now, we are going to prove that any $S_{gf}$ a polynomial of
 positive degree in the Lagrange multipliers $\Lambda =
 (\bar{\lambda}, \bar{\mu}, \ell,m)$ which is both $W$ and $\Stop$
 invariant is of the form
 \be
 S_{gf} = \Stop W X \cdot
 \label{77}
 \ee
 
 Using the homotopy $k_w$ associated with the abelian Weil algebra,
 cf. Eq. (\ref{eq:7}), we can write
 \be
 S_{gf} = (k_W W_\Lambda + W_\Lambda k_W) S_{gf}
 \label{eq:78}
 \ee
 where $W_\Lambda$ is the part of $W$ acting on $\Lambda$. Writing
 \be
 W=W_\Lambda + W_0
 \label{eq:79}
 \ee
 and using 
 \be
 W S_{gf} =0,
 \label{eq:80}
 \ee
 we get
 \bea
 S_{gf} &=& (-k_W W_0 + W_\Lambda k_W) S_{gf} = (W_0 + W_\Lambda)
 k_W  S_{gf} \nonumber \\
 &=& W k_W S_{gf} \cdot
 \label{eq:81}
 \ena
 
 We now construct a homotopy for $\Stop$, extending the homotopoy of
 the Weil algebra: For all pairs $x,y$ such that
 \bea
 \Stop x &=& y + \ell_\om x \nonumber \\
 \Stop y &=& - \ell_{\Om} x + \ell_\om y
 \label{eq:82}
 \ena
 we introduce families $x_t$ such that
 \bea
 \Stop x_t &=& y_t + \ell_{\om_t} \; \; x_t \nonumber \\
 \Stop y_t &=& - \ell_{\Om_t} \; \; x_t + \ell_{\om_t} \; \; y_t
 \label{eq:83}
 \ena
 and we define $k_s (t)$
 \bea
 k_s(t) x_t &=& 0 \nonumber \\
 k_s(t) y_t &=& d_t x_t
 \label{eq:84}
 \ena
 $k_s$ is then defined analogously to Eq.(\ref{eq:7}) as
 \be
 k_s = \int_0^1 k_s(t) \var_t \cdot
 \label{eq:85}
 \ee
 
 Then using
 \be
 \Stop S_{gf} =0 \cdot
 \label{eq:86}
 \ee
 we get
 \be
 S_{gf} = \Stop k_s S_{gf}
 \label{eq:87}
 \ee
 
 Hence
 \be
 S_{gf} = W k_W \Stop k_s S_{gf} \cdot
 \label{eq:88}
 \ee
 
 It is straightforward to check that $k_W$ and $\Stop$ anticommute,
 hence
 \bea
 S_{gf} &=& \Stop W k_W k_s S_{gf} \nonumber \\
 &=& \Stop W \chi
 \label{eq:89}
 \ena
 where $\chi$ has to have bigrading (-1,-1) and is thus of the
 form
 \be
 \chi = <\bar{\mu} , g(a)> + (\bar{\mu}, \ell) \cdot
 \label{eq:90}
 \ee
 
 This gives
 \bea
 S_{gf} &=& \Stop \left( <m, g(a)>+<\bar{\mu}, \frac{\delta g}{\delta
 a}  D_a \lambda> + (m, \var(\ell)) \right) \nonumber \\
 &=& -<\ell, g(a)> - <m, \frac{\delta g}{\delta a} (\psi - D_a
 \om)>  \nonumber \\
 && + <\bar{\lambda}, \frac{\delta g}{\delta a} D_a  \lambda> +
 <\bar{\mu}, \frac{\delta g}{\delta a} D_a \mu> + \nonumber \\
 && <\bar{\mu},  \frac{\delta m}{\delta \und{a}} \cdot \psi - 
 \und{D_a} w \cdot \lambda> \cdot
 - (\ell, \var(l))
 \label{eq:91}
 \ena
 
 There remains to prove that locally over orbit space this provides
 a gauge fixing. So, we have to look at
 \be
 \int \cd \lambda \cd \mu \cd \bar{\lambda}\cd \bar{\mu} \cd \ell
 \cd  m  \; \; e^{S_{gf}}  \cdot \label{eq:92}
 \ee
 
 First, against an integrand of top degree in $\psi$ the $\psi$
 dependence of $S_{gf}$ can be forgotten. Secondly the $m$
 integration insures that the second term yields $\wedge \m \om$, of
 maximal degree in $\om$. The term before last can thus be
 forgotten, and the $\bar{\lambda} \lambda \; \; \bar{\mu} \mu$
 integrations yields twice $det \; \m$ with opposite powers. As
 remarked  earlier one could generalize this to
 \be
 \chi = \bar{\mu} g(a, \dot{a}) + \bar{\mu} \var (\ell, \dot{a}) \cdot
 \label{eq:93}
 \ee
 
 So, we have constructed a complicated integral representation of
 the standard gauge fixing form whose advantage is to allow
 constraining the ultraviolet ambiguities within a tight algebraic
 set up.
 
 This concludes the third exercise.

 \sect{CONCLUSION}
 
 Equivariant cohomology provides an algebraic equipment quite well
 adapted to semi-classical formulations of gauge theories. Some
 integral representations associated with those have been collected
 in these notes, with emphasis on their algebraic structures and
 some of the freedom that is allowed. Similar algebraic techniques
 have been used recently to establish other integral representations\cite{21}
  and concrete formulae\cite{22}. Gauge fixing appears
as a necessity when equivariance with respect to non compact groups comes
into play, each time one is writing an integral representation over
field space as opposed to orbit space. It is in general this way ultraviolet
problems have been so far mastered thanks to locality. It is
suggested that more general gauge fixing procedures than the
conventional ones have to be used in order to respect the geometry.
The question remains open whether, at the expense of introducing
extra local fields, some can be found which respect both locality
and renormalizability.\\
    
\newpage

\noindent{\bf ACKNOWLEDGEMENTS}

\indent

These notes have benefited from numerous discussions with M.~Bauer,
C.~Becchi, G.~Girardi, C.~Imbimbo, F.~Thuillier, J.~Zinn-Justin,
R.~Zucchini to whom the author expresses all his gratitude.

\end{document}